# Betweenness and Diversity in Journal Citation Networks as Measures of Interdisciplinarity

# — A Tribute to Eugene Garfield —

Loet Leydesdorff,*[1] Caroline S. Wagner,[2] and Lutz Bornmann[3]

**Abstract**

Journals were central to Eugene Garfield's research interests. Among other things, journals are considered as units of analysis for bibliographic databases such as the Web of Science (WoS) and Scopus. In addition to disciplinary classifications of journals, journal citation patterns span networks across boundaries to variable extents. Using betweenness centrality (BC) and diversity, we elaborate on the question of how to distinguish and rank journals in terms of interdisciplinarity. Interdisciplinarity, however, is difficult to operationalize in the absence of an operational definition of disciplines; the diversity of a unit of analysis is sample-dependent. BC can be considered as a measure of multi-disciplinarity. Diversity of co-citation in a citing document has been considered as an indicator of knowledge integration, but an author can also generate trans-disciplinary—that is, non-disciplined—variation by citing sources from other disciplines. Diversity in the bibliographic coupling among citing documents can analogously be considered as diffusion of knowledge across disciplines. Because the citation networks in the cited direction reflect both structure and variation, diversity in this direction is perhaps the best available measure of interdisciplinarity at the journal level. Furthermore, diversity is based on a summation and can therefore be decomposed; differences among (sub)sets can be tested for statistical significance. In an appendix, a general-purpose routine for measuring diversity in networks is provided.

[1] * corresponding author; Amsterdam School of Communication Research (ASCoR), University of Amsterdam PO Box 15793, 1001 NG Amsterdam, The Netherlands; loet@leydesdorff.net
[2] John Glenn College of Public Affairs, The Ohio State University, Columbus, Ohio, USA, 43210; wagner.911@osu.edu
[3] Division for Science and Innovation Studies, Administrative Headquarters of the Max Planck Society, Hofgartenstr. 8, 80539 Munich, Germany; bornmann@gv.mpg.de

> *"As a communications system, the network of journals that play a paramount role in the exchange of scientific and technical information is little understood…."*
>
> Garfield (1972, p. 471)

**1. Introduction**

The journal network and its role in collecting and communicating advances in science continues to be a source of debate and challenge to understanding. The network illustrates Herbert Simon's (1962) argument that systems are shaped in hierarchies in order to deal with complexity (Boyack *et al.*, 2014). The journal structures provide order and improve the efficiency of the search for new information. Eugene Garfield enhanced this role by creating additional categories for the evaluation of journals (Bensman, 2007).

Although citations are paper-specific (Waltman & van Eck, 2012), Garfield (1972) constructed the *Science Citation Index* (SCI) and its derivatives (such as the *Social Science Citation Index*) at the journal level (Garfield, 1971). By aggregating citations at that level, one obtains a systems view of the disciplines as they are linked to the subjects covered by the respective journals (Narin, Carpenter, & Berlt, 1972). The Institute of Scientific Information (ISI) developed a journal classification system—the so-called "Web-of-Science subject categories" (WC)—that is often used in scientometric evaluations. Three decades later, however, Pudovkin & Garfield (2002, p. 1113) stated that journals had been assigned to these categories by "subjective and heuristic methods" that did not sufficiently appreciate, or perhaps allow for the visibility of, the relatedness of journals across boundaries (Leydesdorff & Bornmann, 2016). As boundaries are drawn to enhance efficiency, new developments, especially those that bring together disparate ideas in original ways (Uzzi *et al*., 2013), can be disadvantaged by a scheme that relies on incremental additions to the conventional subject categories (Rafols *et al*., 2012).

The categorization of research in terms of disciplines has often been commented upon in the history of science. For example, Bernal (1939 at p. 78) noted that: "From their very nature there must be a certain amount of overlapping…." In 1972, the Organization for Economic Cooperation and Development (OECD, 1972) proposed a systematization of the distinctions between multi-, pluri-, inter-, and trans-disciplinarity as categories for research and higher education (Klein, 2010; Stokols *et al*., 2003) "Multidisciplinary" is used for juxtaposing disciplinary/professional perspectives, which retain separate voices; "interdisciplinary" integrates disciplines; and "transdisciplinary" synthesizes disciplines into larger frameworks (Gibbons *et al*., 1994). We adopt these definitions in this study.

We return to a question which Leydesdorff & Rafols (2011) raised, but did not answer conclusively at the time, namely, how to distinguish and if possible rank journals in terms of their "interdisciplinarity" in such a way as to identify where creative combinations of information are indicated. In this previous study we used the 8,707 journals included in the Journal Citation Reports (JCR) 2008, and explored a number of measures of interdisciplinarity and diversity, also detailed in Wagner *et al*. (2011). In this study, we build on the statistical decomposition of the JCR data 2015 (11,365 journals) using VOSviewer (Leydesdorff *et al*., 2017). A statistical decomposition, however, does not have to be semantically meaningful (Rafols & Leydesdorff, 2009). The advantage of this approach, however, is that the two

problems—of decomposition and interdisciplinarity—can be separated analytically. Furthermore, we exploit advances made in recent years:

1. At the time, we did not sufficiently distinguish between cosine-normalization of the data as more or less standard in the scientometric tradition (Ahlgren *et al*., 2003; Salton & McGill, 1983) and the use of graph-analytical measures such as betweenness centrality that presume binary networks. The distance measures in the two topologies, however, are very different: graph-analytically, one can distinguish shortest paths in the network of *relations*; but the vector space is spanned in terms of *correlations*—that is, including non-relations. Proximity can be expressed in this topology, for example, as a cosine value; and distance accordingly as (1 – cosine).
2. Betweenness centrality—the relative number of times that a node is part of the shortest distance ("geodesic") between other nodes in a network—is an obvious candidate for the measurement of interdisciplinarity; it scored as best for this purpose in the previous comparison. Leydesdorff & Rafols (2011, at p. 93) noted that weighted betweenness could be further explored using the citation values of the links as the weights; but at that time, this concept was still under development (Brandes, 2001; Newman, 2004) and not yet implemented for larger-sized matrices. In the meantime, Brandes (2008) comprehensively discussed betweenness and the measure for valued networks was implemented, for example, in the software package *visone* (available at http://visone.info).
3. Diversity measures have been further developed into "true" diversity measures by Zhang *et al.* (2016); "true" diversity can be scaled at the ratio level so that one can consider percentages in increase or decrease of diversity (Jost, 2006; Rousseau, Zhang, & Hu, in preparation). Furthermore, Cassi *et al*. (2014) proposed that diversity can be decomposed into within-group and between-group diversity. Using a general approximation method for distributions, these authors have developed benchmarks for institutional interdisciplinarity (see also the further elaboration in Cassi *et al*., 2017; cf. Chiu & Chao, 2014). However, we shall argue that one can decompose diversity in terms of the cell values of $(p_i p_j d_{ij})$ because it is a summation. Differences among aggregates of these values can be tested for statistical significance using ANOVA with Bonferroni correction *ex post* (e.g., the Tukey test).

4. The availability of virtually unlimited memory resources using a 64-bit operating system and the further development of software for network analysis (Gephi, ORA, Pajek, UCInet, *visone*, VOSviewer, etc.) enables us to address questions that were previously out of reach.

*1.1. "Interdiscipinarity"*

Interdisciplinarity has remained a fluid concept fulfilling various functions at the interfaces between political and scientific discourse (Wagner *et al*., 2011). Funding agencies and policy makers call for interdisciplinarity from a normative perspective based upon their expectation that boundary-spanning produces creative outputs and can contribute to solving practical problems. For example, in 2015 *Nature* devoted a special issue to interdisciplinarity, stating in the editorial that scientists and social scientists "must work together … to solve grand challenges facing society—energy, water, climate, food, health." On this occasion, van Noorden (2015) collected a number of indicators of interdisciplinarity showing a mixed, albeit optimistic picture; interdisciplinary research, by some measures, has been on the rise since the 1980s. According to this study, interdisciplinary research would have long-term impact (> 10 years) more frequently than disciplinary research (p. 306). Asian countries were shown to publish interdisciplinary papers more frequently than western countries (p. 307).

On the basis of a topic model, Nichols (2014, at p. 747) concluded that 89% of the portfolio of the Directorate for Social, Behavioral, and Economic Sciences (SBE) of the US National Science Foundation is "comprised of IDR (interdisciplinary research)—with 55% of the portfolio identified as having external interdisciplinarity and 34% of the portfolio comprised of awards with internal interdisciplinarity. When dollar amounts are taken into account, 93% of this portfolio is comprised of IDR (…)." Although this result may be partly an effect of the methods used (Leydesdorff & Nerghes, 2017), these impressive percentages show, in our opinion, the responsiveness of (social) scientists to calls for interdisciplinarity by funding agencies.

Is this commitment also reflected in the output of research? Do scientists relabel their research for the purpose of obtaining funds (Mutz *et al*., 2015)? On the output side, the journal literature can be considered as a selection environment at the global level. However, the journal literature has recently witnessed important changes in its orientation toward "interdisciplinarity." Using a new business model, *PLOS ONE* was introduced in 2006 with the objective to cover research from all fields of science without disciplinary criteria. "*PLOS ONE* only verifies whether experiments and data analysis were conducted rigorously, and leaves it to the scientific community to ascertain importance, post publication, through debate and comment" (https://en.wikipedia.org/wiki/PLOS_ONE; MacCallum, 2006, 2011). Although this model is multi-disciplinary, it creates room for the evolution of new standards at the edges of existing disciplines.

It remains difficult to define interdisciplinarity, when disciplines cannot be demarcated clearly. Most if not all of science is a process of seeking diverse inputs in order to create innovative insights. Labels as to whether the results of research are classified as "chemistry" or "physics" are added afterwards. Yet, such classifications structure expectations, behavior, and action. A physicist hired in a medical faculty, for example, has to fulfill a different range of expectations and thus faces another range of options than his/her colleague in a physics department.

The journal literature is mostly structured in terms of specialties because its main function has been to control quality, particularly in the case of specialized contributions. The launch of a new journal and its incorporation into the quality control system of the relevant neighboring journals and databases (including the bibliometric ones) provide practicing scientists with new options. The emergence of a new specialty is often associated with the clustering of journals supporting new developments at the field level (e.g., Leydesdorff & Goldstone, 2014; van den Besselaar & Leydesdorff, 1996). However, the demarcation of disciplines in terms of journals has remained a major problem. As noted, one uses WCs in scientometrics as a proxy, but this generates error (Leydesdorff, 2006).

Interdisciplinarity can also be considered as a variable; neither journals nor departments are mono-disciplinary. The system is operational and therefore in flux. But how could one measure the interdisciplinarity of a journal, a department, or even an individual scholar "while a storm is raging at sea" (Neurath, 1932/33)? Is physical chemistry more or less 'interdisciplinary' than biochemistry? Or are both parts of chemistry? Does it matter when a laboratory for biochemistry is relabeled as molecular biology, and thereafter defined as biology?

When interviewed, for example, physicists who attended the emergence of nanotechnology and nanoscience during the 1990s considered the nano domain as just another domain in physics, whereas material scientists experienced this same development as revolutionary (Wagner, Horlings, Whetsell, Mattsson, & Nordqvist, 2015). A new set of research questions became possible and new journals emerged at relevant interfaces, while existing journals changed their orientations; for example, in terms of what is admissible as a contribution. The material scientists considered nanotechnology and nanoscience as a new discipline, while the physicists did not. As Klein (2010) states: "It is two disciplines, one might say, divided by a common subject" (p. 79).

Unlike hierarchical classifications, a network representation of the relations among disciplines and specialties provides room for operational definitions of interdisciplinarity and measurement. Dense areas in the networks can overlap into areas that are less dense; new densities can emerge in the less dense areas because of recursive interactions; densities at interfaces can be approached from different angles, and then other characteristics may prevail in the perception and hence categorization.

In this study, these larger questions about the (inter)disciplinary dynamics of science are reduced to the seemingly trivial question of the measurement of interdisciplinarity of scholarly journals in a specific year. Can we sharpen the instrument so that an operational definition and measurement of interdisciplinarity become feasible? Using the aggregated journal-journal citation network 2015 based on JCR data, we test two measures which have been suggested for measuring interdisciplinarity. By moving from the top level of "all of science" (11k+ journals) to ten broad fields and then to lower levels of specialties, we hope to be able to say more about the quality of the instruments as well as about the problems of measuring interdisciplinarity. In other words, we entertain two research questions: one substantial, about measuring the interdisciplinarity of journals at different levels, and one methodological, about problems with this measurement.

*1.2    The measurement instruments*

The focus will be on two measures of interdisciplinarity: betweenness centrality and diversity.

1.2.1.   Betweenness centrality

Betweenness centrality (Brandes, 2001, 2008; Freeman, 1978/1979) and its derivatives such as "structural holes" (Burt, 2001) are readily available in software packages as measures for brokering roles between clusters. A high betweenness measure at the node level indicates that the node has a higher than average likelihood of being on the shortest path from one node to another. This position may enable the agent at the node to control the flow between other vertices (Brandes, 2008, at p. 137). Investigators with high betweenness are, for example, better positioned to relay (or withhold) information between research groups (Freeman, 1977; Abbasi, Hossain, & Leydesdorff, 2012). They are advantaged in terms of search.

Algorithms for variants of betweenness centrality were notably implemented in the software package *visone*. Among these variants is the possibility to use weighted networks (Freeman *et al.*, 1991). The combination of betweenness centrality with the disparity notion in diversity studies—to be discussed below—is also the subject of ongoing research on Q- or Gefura measures (Flom, Friedman, Strauss, & Neaigus, 2004; Rousseau & Zhang, 2008; Guns & Rousseau, 2015). However, this further extension is not studied here.

1.2.2. Diversity

Following a series of empirical studies by Alan Porter and his colleagues (Porter *et al.*, 2006, 2007, 2008, and 2009), on the one hand, and Stirling's (2007) mathematical elaboration, on the other, Rafols & Meyer (2010) distinguished three aspects of interdisciplinarity: (1) variety, (2) balance, and (3) disparity. Variety can be measured, for example, as Shannon entropy. The participating disciplines in specific instances of interdisciplinarity can be assessed in terms of their balance: a balanced participation can be associated with interdisciplinarity, whereas an unbalanced one suggests a different relationship (Nijssen, Rousseau, & Van Hecke, 1998). In the extreme case, the one discipline is enrolled by the other in a service relationship.

On the basis of animations of newly emerging journal structures, Leydesdorff & Schank (2008) showed that interdisciplinary developments occur often at specific interfaces between disciplines, but are initially presented as—and believed to be—interdisciplinary. From this perspective, interdisciplinarity can be associated with the idea of a pre-paradigmatic phase in the development of disciplines and specialties (van den Daele, Krohn, & Weingart, 1979). New and initially interdisciplinary developments may crystallize into new disciplinary structures (van den Besselaar & Leydesdorff, 1996) or they may dissipate as the core disciplines absorb the new concepts.

The measurement of "disparity" provides us with an ecological perspective: a collaboration between authors in biology and chemistry, for example, can be considered as less interdisciplinary in terms of disparity than one between authors in chemistry and anthropology. The cognitive distance between latter two disciplines—being a natural and a social science discipline is much larger than that between two neighboring fields in the natural sciences (Boschma, 2005). The disparity thus reflects a next-order structure in terms of ecological distances and niches among journal sets.

Disparity and variety can be combined in the noted measures of diversity (Rao, 1982; Stirling, 2007) as follows:

$$\Delta = \sum_{\substack{i,j \\ (i \ne j)}} p_i p_j d_{ij} \qquad (1)$$

In this formula, $i$ and $j$ represent different categories; $p_i$ represents the relative frequency or probability of category $i$, and $d_{ij}$ the distance between $i$ and $j$. The distance, for example, can be the geodesic (that is, shortest path) in a network or (1 – cosine) in a vector space by using the cosine as a proximity measure (Ahlgren *et al*., 2003; Jaffe, 1989; Salton & McGill, 1983). The multiplication of the measures of distance and relative occupation has led to the characterization of this measure as "quadratic entropy" (e.g., Izsáki & Papp, 1995). Stirling (2007) suggests developing a further heuristics by weighing the two components; for example, by adding exponents. However, one then obtains a parameter space which is infinite (Ricotta & Szeidl, 2006).

The first part of Equation 1 (that is, the measure of variety $\sum_{\substack{i,j \\ (i \neq j)}} p_i p_j$) is also known as the Gini-Simpson diversity measure in biology or the Herfindahl-Hirschman index in economics (Leydesdorff, 2015). Note that this term is measured at the level of a vector. Using a citation matrix, two different distance matrices can be constructed among the citing and cited vectors, respectively. In the citing dimension, Rao-Stirling diversity has been considered as a measure of integration in interdisciplinary research (Porter & Rafols, 2009; Rousseau *et al.*, in preparation; Wagner *et al.*, 2011, p. 16). Variety and disparity are combined and integrated in a citing paper by the citing author(s). In the cited dimension, one measures diversity (in terms of variety and disparity) in the structures from which one cites. The structures operate as selection environments. Rousseau *et al.* (2017) suggests that diversity in the cited dimension should be considered as diffusion: diffusion can be interdisciplinary to various extents.

Zhang, Rousseau, & Glänzel (2016) further developed $\Delta$ into $^2D^3$ as a "true" diversity measure; true diversity has the advantage that the measure is scaled so that a 20% higher value of $^2D^3$ indicates 20% more diversity (Jost, 2006). Conveniently, the two measures relate monotonically as follows (Zhang *et al.*, 2016, p. 1260, Eq. 6)*:*

$$^2D^3 = 1/(1 - \Delta) \qquad (2)$$

True diversity varies from one to infinity when $\varDelta$ varies between zero and one. Note that these diversity measures do not include "balance" as the third element distinguished in the definition of interdisciplinarity by Rafols & Meyer (2010). One can envisage adding a third probability distribution ($p_k$) to Eq. 1 as a representation of the disciplinary contributions (Rafols, 2014). Alternatively, "balance" can be operationalized using, for example, the Gini-index.

As noted, Cassi *et al.*, (2014) developed a methodology for the decomposition of diversity into within-group and between-group diversity (see also the further elaboration in Cassi *et al.*, 2017; cf. Chiu & Chao, 2014). In our opinion, the equations are valid for each subset since the operation is a straightforward summation. Consequently, one can decompose diversity in terms of the cell values ($p_i p_j d_{ij}$). Differences among aggregated subsets can be tested using ANOVA with Bonferroni correction *ex post* (e.g., the Tukey test). For the exploration of this decomposition, we use the WoS Category Library and Information Science (86 journals in the JCR 2015) instead of the set of 62 journals categorized by Leydesdorff *et al.* (2017) into a single group on statistical grounds. The results are then easier to follow.

*1.3. Units of analysis*

In addition to the various aspects of interdisciplinarity that can be distinguished, the choice of the system of reference will make a difference. Interdisciplinarity can be attributed to departments, journals, œuvres, emerging disciplines, etc. In science studies, it is customary to distinguish between the socially organized group level and the level of intellectually organized fields of science (Whitley, 1984). The interdisciplinarity of a group (e.g., a department) can be important from the perspective of team science. The dynamics of interdisciplinary at the field level are relatively autonomous ("self-organizing"; van den Daele & Weingart, 1975).

For example, the interdisciplinary development of nano-technology in the 1990s required contributions from chemistry (e.g., advanced ceramics), applied physics, and materials sciences. A group in a chemistry faculty will be positioned for the challenge of participating in this new development differently from a group in physics. The group dynamics, in other words, can be different among groups and from the field dynamics. New fields of science may develop at the global level, whereas groups are localized. One can also consider the fields as the selection environments for groups or, more generally, individual or institutional agency. Selection mechanisms can reflexively be anticipated.

Furthermore, one can attribute interdisciplinarity as a variable to units of analysis such as authors, groups, texts at the nodes of networks, or second-order units of analysis such as links. Factor loadings, for example, are attributes of variables. One can expect different dynamics at the first-order or second-order level. Whereas interdisciplinarity can a political or managerial objective in the case of first-order units (e.g., groups), interdisciplinarity attributed at the level of second-order units (e.g., fields) is largely beyond the control of decision makers or individual scientists. Second-order units can be rearranged and thus develop resilience against external steering.

Note that these distinctions are analytical: journals, for example, are organized in terms of their production process, but can be self-organizing in terms of their content to variable extents. The interdisciplinarity of a journal or a department is also determined by the sample and the level of granularity in the analysis. A journal, for example, may appear interdisciplinary in the context of a large set of journals, but when this set is decomposed, the interdisciplinarity may be lost since the borders are drawn differently. For example, important ties to other domains may be cut by decomposition. In sum, one unavoidably entertains a model when measuring "interdisciplinarity;" and by using this model, the concept is (re)constructed.

## 2. Methods

*2.1. Data*

We use the directed (asymmetrical and valued) 1-mode matrix among the 11,365 journals listed in the Science and Social Sciences Citation Index in 2015. Table 1 provides descriptive statistics of the largest component of 11,359 journals. (Six journals are not connected.)[4]

**Table 1**: Network characteristics of the largest component of the matrix based on JCR 2015.

|  | *JCR 2015* |
|---|---|
| **N of journals (nodes)** | 11,359 |
| **Links** | 2,848,736 |
|  | (11,049 loops) |
| **Total citations** | 43,010,234 |
| **Density** | 0.022 |
| **Average (total) degree** | 501.582 |
| **Cluster coefficient** | 0.220 |
| **Avg. distance** | 2.495 |
| **Maximum distance** | 6 |

---

[4] *Avian Res, EDN, Neuroforum, Austrian Hist Yearb, Curric Matters,* and *Policy Rev.*

In the first round of the decomposition, ten clusters were distinguished. These are listed in Table 2. At http://www.leydesdorff.net/jcr15/scope/index.htm the reader will find a hierarchical decomposition in terms of maps of science.

**Table 2**: Fields distinguished at the top level of JCR 2015.

|    | *Field-designation* | N |
|----|---------------------|-------|
| 1  | Social Sciences | 3,274 |
| 2  | Computer Science | 2,003 |
| 3  | Medicine | 1,965 |
| 4  | Environmental Sci | 1,595 |
| 5  | Biomedical | 784 |
| 6  | Chemistry | 652 |
| 7  | Bio-agricultural | 583 |
| 8  | Physics | 440 |
| 9  | Ophthalmology | 57 |
| 10 | Data analysis ("Big data") | 6 |
|    | | 11,359 |

We pursue the analysis for the complete set (n = 11,359) and for the first cluster (n = 3,274). Within this latter subset, 62 journals are classified as Library and Information Science (LIS) in the second round of decomposition. We use the LIS set as an example at the (next-lower) specialty level.[5]

*2.2. Statistics*

As noted, we focus in this study on betweenness centrality and diversity as two main candidates for measuring interdisciplinarity in journal citation networks. The betweenness centrality (BC) of a vertex $k$ is equal to the proportion of all the geodesics between pairs ($g_{ij}$) of vertices that include this vertex ($g_{ijk}$; e.g., de Nooy *et al.*, 2011, p. 151). The BC for a vertex *k* can formally be written as follows:

$$BC_k = \sum_i \sum_j \frac{g_{ijk}}{g_{ij}}, i \neq j \neq k \tag{3}$$

---

[5] For the delineation of the LIS set, see the appendix of Leydesdorff *et al.* (2017) at pp. 1611f.

Freeman (1977) introduced several variants of this betweenness measure when proposing it. In their study of centrality in valued graphs, Freeman, Borgatti, & White (1991) further elaborated flow centrality, which includes all the independent paths contributing to BC in addition to the geodesics. In the meantime, a number of software programs for network analysis have adopted Brandes' (2008) algorithm for valued graphs.[6] We use Pajek and UCInet for non-valued graphs and *visone* for analyzing valued ones.[7]

While BC can be computed on an asymmetrical matrix, Rao-Stirling diversity and $^2D^3$ are evaluated along vectors in either the cited or citing direction of a citation matrix. Both the proportions and the distances have to be taken in the one direction or the other. One thus obtains two different—but most likely correlated—measures. The proportions are straightforward relative to the sum of the references given by the journal (citing) or the citations received (cited).[8] The distance measure, however, provides us with another parameter choice.

---

[6] Non-valued BC can be obtained by first binarizing the matrix.
[7] UCInet offers also a number of options such as "attribute weighted betweenness centrality," but the results are sometimes very similar to ordinary BC.
[8] The "total citations" provided by the JCR can be considerably larger than the citations included in the matrix. Probably, citations by other sources such as the Arts & Humanities Citation Index are also included. We correct for this by recounting the total citations (and total references) in each set.

In line with the reasoning about BC, one could consider using geodesics as a measure of distance. However, the average geodesic in the network under study is 2.5 with a standard deviation of 0.6 (Table 1). In other words, the variation in the geodesic distances is small: most of them are 2 or 3.[9] The choice of another distance measure—or equivalently (1 – proximity)—provides us with a plethora of options. We chose (1 – cosine) as the distance measure because Euclidean distances did not work in our previous project (Leydesdorff & Rafols, 2011). The cosine has been used as a proximity measure in technology studies by Jaffe (1989); Ahlgren, Jarneving, & Rousseau (2003) suggested using the cosine (Salton & McGill, 1983) as an alternative to the Pearson correlation in bibliometrics.

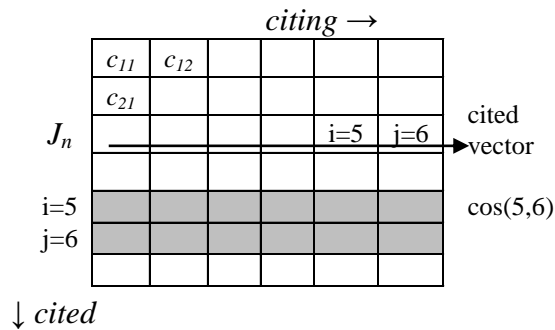

**Figure 1**: The computation of Rao-Stirling diversity.

The computation is intensive because of the permutation of the $i$ and $j$ parameters along the vector in each case. A routine for generating diversity values on the basis of a Pajek file is provided in the Appendix. Note that $p_i$ and $p_j$ along a vector can both be larger than zero, but the cosine between the vectors $i$ and $j$ in the same direction may be zero. For example, if a journal $n$ is cited by two marginal journals ($i = 5$ and $j = 6$ in Figure 1), the co-occurrence in the vertical direction is larger than zero; but in the horizontal direction the cosine value can be zero and the distance therefore one. Since the probabilities are by definition normalized for size, the cosine values between marginal journals may thus boost diversity as measured here. (Given the skew in scientometric distributions, one can expect relative marginality to prevail in any delineated domain.)

---

[9] A transformation to a measure between zero and one could be (1 – 1/N) leading to a distance of ½ in the case of a distance of two and 0.67 in the case of three.

## 3. Results

*3.1. The full set of 11,365 journals in JCR 2015*

Table 3 lists the top-25 journals when ranked for betweenness, valued betweenness, and diversity measured as $^2D^3$ in both the citing and cited directions. Not surprisingly, *PLOS ONE* ranks highest on BC in both the binary and the valued case. The Pearson correlation for the two rankings across the file is larger than .99 (Table 4), but differences at the top of the list are sometimes considerable. *PLOS ONE* and, for example, *Psychol Bull* gain in score when BC is based on values, but *Nature* and *Science* lose.

Interestingly, *Scientometrics* ranks 11$^{th}$ on BC in the binary case, but only in the 15$^{th}$ place using valued BC. Typically, this journal cites and is cited by journals in other fields incidentally and unsystematically in addition to more dense citation in its own intellectual environment. Citations to and from *Psychol Bull* in contrast are more specific. *Annu Rev Psychol* and *Psychol Rev* show the same pattern as *Psychol Bull* of increasing BC when valued.

**Table 3**: Top 25 journals in terms of various betweenness centrality and diversity measures.

| Journal | BC (Pajek) | Journal | Valued % BC | Journal | Diversity $^2D^3$ Cited | Journal | Diversity $^2D^3$ Citing |
|---|---|---|---|---|---|---|---|
| PLOS ONE | 16.56 | PLOS ONE | 17.30 | Am Behav Sci | 17.56 | J Chin Inst Eng | 20.13 |
| P Natl Acad Sci Usa | 5.11 | P Natl Acad Sci Usa | 4.97 | Daedalus-Us | 16.67 | Sci Iran | 16.22 |
| Soc Sci Med | 3.40 | Soc Sci Med | 3.30 | Ann Am Acad Polit Ss | 16.06 | Arab J Sci Eng | 15.89 |
| Sci Rep-Uk | 2.85 | Psychol Bull | 2.06 | P Ieee | 15.2 | Teh Vjesn | 15.83 |
| Nature | 2.26 | Sci Rep-Uk | 1.71 | Field Method | 14.58 | J Cent South Univ | 15.4 |
| Science | 2.12 | Am J Public Health | 1.69 | Qual Quant | 14.25 | Adv Mech Eng | 14.85 |
| Am J Public Health | 1.75 | Nature | 1.61 | Am J Econ Sociol | 13.73 | J Mar Sci Tech-Taiw | 14.56 |
| Psychol Bull | 1.50 | Science | 1.47 | Am J Sociol | 13.57 | J Test Eval | 14.35 |
| Energ Policy | 1.19 | Energ Policy | 0.96 | Brit J Sociol | 13.39 | Measurement | 14.29 |
| Ecol Econ | 0.95 | Ecol Econ | 0.87 | Technol Rev | 13.11 | Sustainability-Basel | 14.1 |
| Scientometrics | 0.85 | Annu Rev Psychol | 0.84 | Annu Rev Sociol | 12.92 | Dyna-Bilbao | 14.03 |
| Nat Commun | 0.78 | Psychol Rev | 0.61 | Philos T R Soc A | 12.84 | J Zhejiang Univ-Sc A | 13.99 |
| Sustainability-Basel | 0.78 | Manage Sci | 0.60 | Crit Inquiry | 12.49 | Math Probl Eng | 13.78 |
| Manage Sci | 0.71 | Phys Rev E | 0.59 | Am Hist Rev | 12.37 | J Eng Res-Kuwait | 13.32 |
| Phys Rev E | 0.66 | Scientometrics | 0.57 | Am Sociol Rev | 12.18 | Sadhana-Acad P Eng S | 12.92 |
| Biomed Res Int | 0.59 | Global Environ Chang | 0.48 | P Roy Soc A-Math Phy | 11.99 | Sains Malays | 12.81 |
| Global Environ Chang | 0.58 | Sensors-Basel | 0.45 | Curr Sociol | 11.88 | Ieee Lat Am T | 12.48 |
| Annu Rev Psychol | 0.58 | Trends Cogn Sci | 0.44 | Risk Anal | 11.85 | Qual Quant | 12.44 |
| Sensors-Basel | 0.57 | Environ Sci Technol | 0.42 | Harvard Bus Rev | 11.67 | Ksce J Civ Eng | 12.15 |
| Sci Total Environ | 0.53 | Phys Rev Lett | 0.40 | Struct Equ Modeling | 11.65 | Front Inform Tech El | 12.1 |
| Environ Sci Technol | 0.51 | J Affect Disorders | 0.39 | Foreign Aff | 11.62 | J Sensors | 11.95 |
| Comput Educ | 0.47 | J Geophys Res | 0.38 | Am Psychol | 11.49 | Natl Acad Sci Lett | 11.74 |
| Psychol Rev | 0.46 | Expert Syst Appl | 0.37 | Comput J | 11.41 | B Pol Acad Sci-Tech | 11.63 |
| Lancet | 0.44 | Appl Math Comput | 0.37 | J Appl Soc Psychol | 11.3 | Rev Estud Soc | 11.55 |
| J Affect Disorders | 0.44 | Nat Commun | 0.35 | J Zhejiang Univ-Sc A | 11.27 | Maejo Int J Sci Tech | 11.47 |

Most of the journals with high BC values are multi-disciplinary journals. In accordance with its definition, BC measures the extent to which the distance between otherwise potentially distant clusters is bridged. Note that some journals in the social sciences score high on BC, among which is *Scientometrics.* In our opinion, *Scientometrics* can be considered as a specialist journal with a specific disciplinary orientation. As noted, however, its citation patterns and being cited patterns span across different disciplines because a variety of disciplines can be the subject of study and indicators are used in other fields. In other words, BC does not teach us about the nature of the knowledge production process, but about patterns of integration and diffusion across disciplinary boundaries (Rousseau *et al*., 2017, in preparation).

**Table 4**: Pearson and Spearman' rank correlations of BC and $^2D^3$ among 11,359 journals (in the lower and upper triangle, respectively); all correlations are significant at the level <.01.

|  | *BC* | *Valued BC* | *$^2D^3$ cited* | *$^2D^3$ citing* |
|---|---|---|---|---|
| *BC* |  | .964 | .396 | .190 |
| *Valued BC* | .993 |  | .472 | .231 |
| *$^2D^3$ cited* | .050 | .041 |  | .442 |
| *$^2D^3$ citing* | .042 | .033 | .406 |  |

Table 4 shows that BC and $^2D^3$ measure different things. Diversity in the citing direction is not correlated to BC. In the cited direction the rank-order correlation is still substantial. This correlation can be explained as follows: the disparity factor ($d_{ij}$) indicates the distances that have to be bridged between different domains. The (multi-disciplinary) structure of science is reflected in both this distance and BC. However, variety [$\sum_{\substack{i,j \\ (i \neq j)}} p_i p_j$]—as the second component of diversity—is based on a different principle. In the citing dimension, particularly, one may cite across disciplinary boundaries ("trans-disciplinarily"; Gibbons *et al*., 1994) and generate variety. This source of variation is also reflected in the cited dimension, since the cited can be considered as the archive of a time-series of citing relations. Not incidentally, therefore, we find journals in the right-most column of Table 3 from the periphery, or with a specific national background that may be problem- or sector-oriented (e.g., agriculture). Leydesdorff & Bihui (2005) found such a non-disciplinary orientation in the case of Chinese journals that are institutionally based.

The *Journal of the Chinese Institute of Engineering* (with $^2D^3$ = 20.13 at the top of this list), for example, was cited in 2015 in articles published in 56 journals, but it cites from 230 journals. It can therefore be considered a net importer of knowledge (Yan, Ding, Cronin, & Leydesdorff, 2013). Figure 2 shows this environment of 230 journals as a map based on aggregated citation relations. Using BC as the values for the nodes, Figure 2A first shows the structure of the journals as a map. In Figure 2B, diversity in the citing direction is used as the parameter for the node sizes. This brings engineering journals more to the fore then physics journals. The *J Chin Inst Eng* itself is not visible in Figure 2A, but most pronouncedly in Figure 2B. Note that there is further nothing special about this journal: its 2-year Journal Impact Factor (JIF) is .246 and the 5-year JIF is .259.

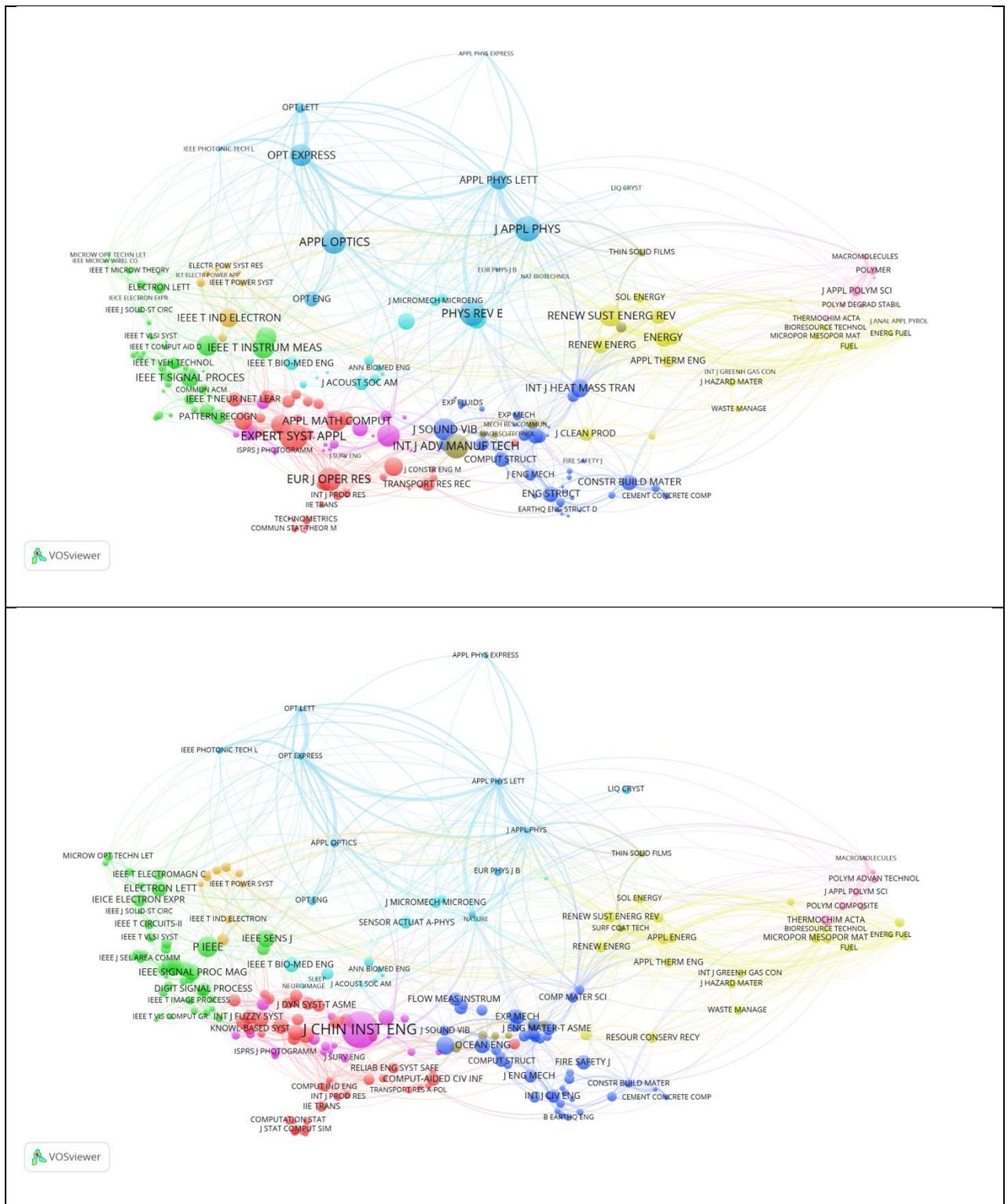

**Figure 2**: Citing patterns among 230 journals cited by *J Chin Inst Eng* during 2016. Nodes are sized according to BC in Figure 2A but according to diversity in the citing dimension in Figure 2B.

*3.2. Journals in the Social Sciences*

The largest subset of journals distinguished in the decomposition (Table 2) is a group of 3,274 journals in the social sciences. We pursue the analysis for this subset in order to see whether the patterns found above can be considered general. In a next section we zoom further in to the subset of journals classified as LIS within this set.

Table 5 shows that *Soc Sci Med* is ranked highest in terms of BC in both the valued and non-valued analysis. This journal was ranked in the third position in the full set—after *PLOS ONE* and *PNAS*. The rank of *Scientometrics* has now decreased from the 11$^{th}$ to the 21$^{st}$ position using non-valued BC and from the 15$^{th}$ to the 31$^{st}$ position using valued BC. A large proportion of its betweenness is in connecting social science disciplines with the natural and medical sciences. These relations across disciplinary divides are cut by the decomposition.

**Table 5**: Top 25 social-science journals (n = 3,274) in terms of various betweenness centrality and diversity measures.

| Journal | BC (Pajek) | Journal | Valued BC (visone) | Journal | $^2D^3$ cited | Journal | $^2D^3$ citing |
|---|---|---|---|---|---|---|---|
| Soc Sci Med | 7.38 | Soc Sci Med | 6.64 | Soc Res | 18.11 | Disasters | 9.15 |
| Am J Public Health | 2.62 | Psychol Bull | 2.18 | Qual Res Psychol | 17.92 | Convergencia | 8.64 |
| Psychol Bull | 2.34 | Am J Public Health | 2.06 | Am Behav Sci | 17.39 | Continuum-J Media Cu | 8.13 |
| Front Psychol | 2.25 | Front Psychol | 1.75 | Daedalus-Us | 16.44 | Evaluation Rev | 7.79 |
| Comput Hum Behav | 2.04 | Comput Hum Behav | 1.61 | Ann Am Acad Polit Ss | 15.94 | Indian J Gend Stud | 7.71 |
| J Bus Ethics | 1.61 | J Pers Soc Psychol | 1.26 | Field Method | 14.97 | Food Cult Soc | 7.3 |
| J Pers Soc Psychol | 1.31 | J Bus Ethics | 1.20 | Qual Inq | 14.91 | Educ Sci-Theor Pract | 7.2 |
| World Dev | 1.24 | World Dev | 1.04 | Nation | 13.81 | Hum Organ | 7.1 |
| Pers Indiv Differ | 1.22 | Pers Indiv Differ | 1.04 | Soc Sci Inform | 13.8 | Etikk Praksis | 7.09 |
| Soc Indic Res | 1.08 | Soc Indic Res | 0.89 | Am J Econ Sociol | 13.56 | Inform Cult | 7.03 |
| Appl Econ | 1.04 | J Adv Nurs | 0.79 | Am J Sociol | 13.38 | China Rev | 7.01 |
| Am Psychol | 0.94 | Am Sociol Rev | 0.78 | Brit J Sociol | 13.31 | Curr Sociol | 6.85 |
| J Adv Nurs | 0.92 | Am Psychol | 0.70 | Theor Soc | 12.95 | Inform Res | 6.84 |
| J Pragmatics | 0.91 | Appl Econ | 0.70 | Qual Res | 12.9 | Educ Assess Eval Acc | 6.83 |
| Am Sociol Rev | 0.90 | Geoforum | 0.66 | Signs | 12.81 | Crit Asian Stud | 6.76 |
| Am Econ Rev | 0.80 | Am J Sociol | 0.65 | Annu Rev Sociol | 12.75 | Hist Soc Res | 6.65 |
| Geoforum | 0.80 | J Pragmatics | 0.63 | J Socio-Econ | 12.75 | Environ Plann C | 6.62 |
| J Bus Res | 0.79 | J Appl Psychol | 0.62 | Crit Inquiry | 12.38 | Eur J Womens Stud | 6.61 |
| Am J Sociol | 0.78 | Am Econ Rev | 0.61 | J R Stat Soc A Stat | 12.38 | Educ Xx1 | 6.53 |
| J Appl Psychol | 0.73 | J Bus Res | 0.58 | Am Hist Rev | 12.36 | Eval Program Plann | 6.49 |
| Scientometrics | 0.73 | J Econ Behav Organ | 0.51 | Sociol Methodol | 12.09 | Econ Soc | 6.47 |
| Psychol Sci | 0.68 | Annu Rev Sociol | 0.50 | Am Sociol Rev | 12.05 | Fem Psychol | 6.47 |
| J Econ Behav Organ | 0.68 | Ecol Econ | 0.49 | Qual Quant | 11.84 | Health Soc Care Comm | 6.44 |
| Curr Anthropol | 0.65 | Cognition | 0.49 | Curr Sociol | 11.81 | Cult Stud | 6.39 |
| Comput Educ | 0.64 | Child Dev | 0.49 | New Left Rev | 11.79 | Child Soc | 6.39 |

**Table 6**: Pearson and Spearman' rank correlations of BC and $^2D^3$ among 3,274 journals (in the lower and upper triangle, respectively).

|  | *BC* | *Valued BC* | *$^2D^3$ cited* | *$^2D^3$ citing* |
|---|---|---|---|---|
| ***BC*** |  | .983** | .466** | .153** |
| ***Valued BC*** | .994** |  | .475** | .176** |
| ***$^2D^3$ cited*** | .194** | .176** |  | .205** |
| ***$^2D^3$ citing*** | .010 | .008 | .120 |  |

\*\*. Correlation is significant at the 0.01 level (2-tailed).
\*. Correlation is significant at the 0.05 level (2-tailed).

The pattern described above for the full set is also found in this subset. Two factors explain 78.6% of the variance in the four variables: BC versus the variation-factor in the diversity citing (Table 7). The Pearson correlations between BC (binary and valued) and $^2D^3$ citing are .010 and .008, respectively. Note the negative sign of $^2D^3$ *citing* on factor 1 in Table 7. The two mechanisms thus stand orthogonally.

**Table 7**: Varimax rotated factor solution for the four variables; $n = 3264$.

**Rotated Component Matrix**[a]

|  | Component 1 | Component 2 |
|---|---|---|
| ***BC*** | .990 |  |
| ***valued BC*** | .990 |  |
| ***$^2D^3$ citing*** | -.109 | .823 |
| ***$^2D^3$ cited*** | .223 | .661 |

Extraction Method: Principal Component Analysis. Rotation Method: Varimax with Kaiser Normalization.

The journals in the right-most column of Table 6 are recognizably trans-disciplinary or, in other words, reaching out across boundaries. On the cited side, the pronounced position of sociology journals is noteworthy. Major sociology journals such as *Am J Sociol*, *Brit J Sociol*, and *Am*

*Sociol Rev* figure in this top list as they did in Table 3, but other sociology journals such as *Soc Sci Inform* also rank high on this list (#9), while this journal was ranked only at position 1,363 in the total set.

In Figure 3, we try to capture the differences visually. Figure 3A first provides a map of these 3,274 journals. In Figure 3B, the node sizes are proportional to the BC scores of the journals. One can see a shift to the applied side. For example, the *Am Rev Econ* comes to the foreground in the left-most cluster (pink) in Figure 3A, while this most-pronounced position is assumed by *Appl Econ* and *World Dev* in Figure 3B. Similarly, *J Pers Soc Psychology*—the flagship of this field—is overshadowed by *Psych Bull* in the top-right cluster (turkois) of Figure 3B. This journal is also read outside the specialty. The *J Bus Ethics* is most pronounced in terms of BC values among the business and management journals in the light-blue cluster top-left.

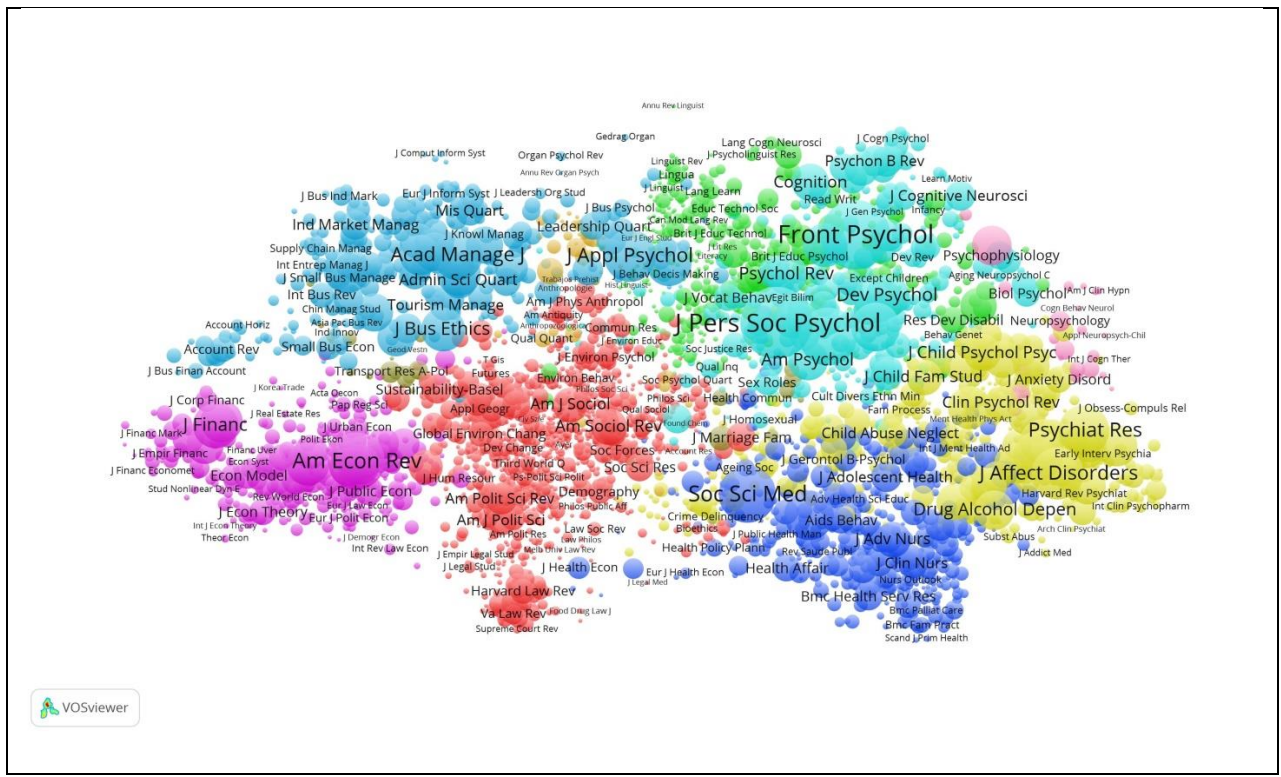

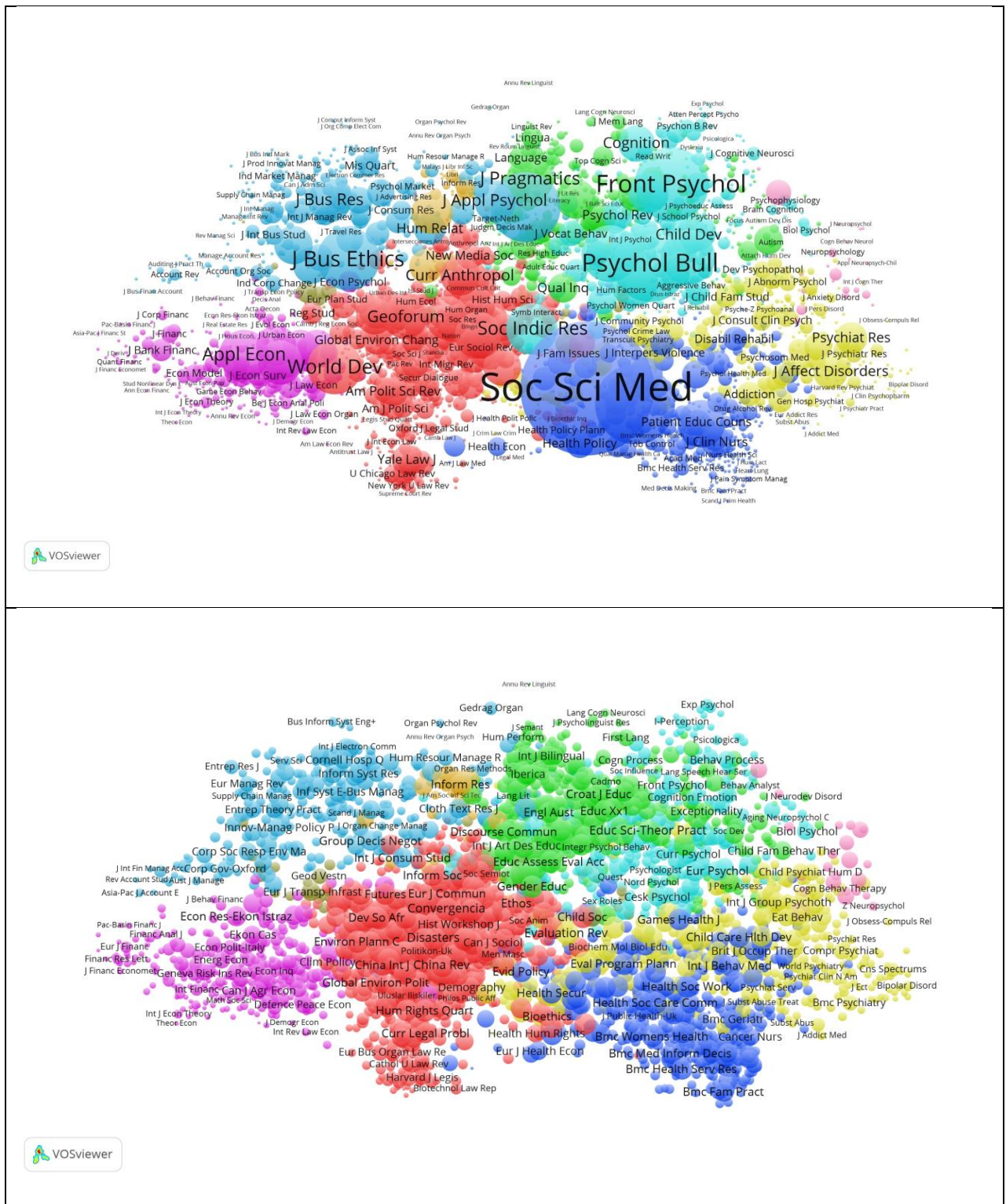

**Figure 3**: Comparison of the map for the social sciences (Fig. 3A) with one using BC (Figure 3B) and $^2D^3$ values citing (Figure 3C) for the sizes of the nodes (in VOSviewer).

In Figure 3C, the node sizes are proportional to the diversity scores (citing). The picture teaches us that highly diverse journals are spread across the disciplines as variation. All disciplines have portfolios of journals of which some are more diverse than others.

*3.3. Library and information sciences (LIS)*

We first pursued the analysis using the 62 journals that were classified as LIS, but for reasons of presentation, here we use the results of the analysis based on citations among the 86 journals classified in terms of SC as LIS in the JCR 2015. Otherwise, the discussion about the differences between the two samples would lead us away from the objectives of this study (cf. Leydesdorff *et al*., 2017).

**Table 8**: Top 25 LIS journals in terms of various betweenness centrality and diversity measures.

| Journal | BC (Pajek) | Journal | Valued BC (visone) | journal | $^2D^3$ cited | Journal | $^2D^3$ citing |
|---|---|---|---|---|---|---|---|
| Scientometrics | 6.82 | Scientometrics | 10.12 | Inform Soc | 4.28 | Libr Inform Sci Res | 3.78 |
| J Doc | 5.79 | Libr Inform Sci Res | 7.38 | J Comput-Mediat Comm | 3.89 | Online Inform Rev | 3.76 |
| Libr Inform Sci Res | 4.36 | J Acad Libr | 7.06 | Online Inform Rev | 3.39 | J Assoc Inf Sci Tech | 3.72 |
| J Acad Libr | 3.82 | J Libr Inf Sci | 7.04 | Int J Inform Manage | 3.32 | Aslib J Inform Manag | 3.69 |
| Electron Libr | 3.71 | J Doc | 6.22 | Aslib Proc | 3.26 | Electron Libr | 3.67 |
| Online Inform Rev | 3.71 | Inform Res | 5.25 | Inform Dev | 3.15 | Inform Res | 3.51 |
| Int J Inform Manage | 3.42 | Mis Quart | 5.05 | Inform Manage-Amster | 3.11 | J Inf Sci | 3.48 |
| Inform Res | 3.41 | Electron Libr | 4.89 | J Inf Sci | 3.10 | Can J Inform Lib Sci | 3.46 |
| Mis Quart | 3.35 | Online Inform Rev | 4.06 | Serials Rev | 2.93 | Investig Bibliotecol | 3.30 |
| Inform Manage-Amster | 2.92 | Gov Inform Q | 3.70 | J Med Libr Assoc | 2.90 | Inform Cult | 3.23 |
| Gov Inform Q | 2.57 | Inform Manage-Amster | 3.22 | Ethics Inf Technol | 2.66 | Libr Hi Tech | 3.10 |
| J Assoc Inf Sci Tech | 2.25 | J Assoc Inf Sci Tech | 3.21 | J Scholarly Publ | 2.57 | J Doc | 3.07 |
| J Inf Sci | 2.19 | Int J Inform Manage | 2.90 | J Inf Technol | 2.47 | Program-Electron Lib | 3.02 |
| Libr Trends | 1.91 | Libr Trends | 2.57 | Learn Publ | 2.46 | Inform Soc-Estud | 2.99 |
| J Manage Inform Syst | 1.72 | J Inf Sci | 2.25 | Libr Hi Tech | 2.44 | J Libr Inf Sci | 2.98 |
| J Libr Inf Sci | 1.69 | Inform Soc | 2.12 | Libr Inform Sci Res | 2.41 | Afr J Libr Arch Info | 2.92 |
| Inform Soc | 1.62 | Inform Process Manag | 1.68 | J Doc | 2.40 | Libri | 2.91 |
| Coll Res Libr | 1.57 | J Manage Inform Syst | 1.60 | Libr Trends | 2.38 | Libr Trends | 2.89 |
| Inform Dev | 1.40 | Inform Dev | 1.54 | Program-Electron Lib | 2.36 | Libr Resour Tech Ser | 2.89 |
| J Informetr | 1.40 | Inform Technol Peopl | 1.47 | Inform Technol Libr | 2.36 | Inform Process Manag | 2.84 |
| Inform Process Manag | 1.37 | Coll Res Libr | 1.31 | J Am Soc Inf Sci Tec | 2.33 | Libr Quart | 2.83 |
| Inform Technol Peopl | 1.23 | J Knowl Manag | 1.12 | Inform Res | 2.33 | Learn Publ | 2.71 |
| J Assoc Inf Syst | 1.05 | J Med Libr Assoc | 0.91 | Aust Acad Res Libr | 2.33 | Malays J Libr Inf Sc | 2.70 |
| Eur J Inform Syst | 1.02 | Telecommun Policy | 0.86 | J Strategic Inf Syst | 2.32 | Inform Soc | 2.60 |
| Libr Hi Tech | 0.99 | Inform Technol Dev | 0.84 | Inform Technol Peopl | 2.31 | Ref User Serv Q | 2.60 |

**Table 9**: Pearson and Spearman' rank correlations of BC and $^2D^3$ among 86 LIS journals; in the lower and upper triangle, respectively.

|  | BC | Valued BC | $^2D^3$ cited | $^2D^3$ citing |
|---|---|---|---|---|
| **BC** |  | .987** | .541** | .416** |
| **Valued BC** | .938** |  | .538** | .422** |
| **$^2D^3$ cited** | .324** | .256* |  | .282** |
| **$^2D^3$ citing** | .329** | .346** | .230* |  |

\*\*. Correlation is significant at the 0.01 level (2-tailed).
\*. Correlation is significant at the 0.05 level (2-tailed).

In Table 8, *Scientometrics* is the journal with the highest BC in both analyses. *JASIST* follows at only the 12th position, while one would expect the latter journal to be more integrative among the different subjects studied in LIS. In terms of knowledge integration indicated as diversity in the citing dimension, *JASIST* assumes the third position and *Scientometrics* trails in 45th position. In the cited dimension, the diversity of *Scientometrics* is ranked 70 (among 86). Thus, the journal is cited in this environment much more specifically than in the larger context of all the journals included in the JCR, where it assumed the 339th and 6,246th position among 11,359 observations, respectively. In the latter case the quantile values are 97.0 and 45.0, respectively, *versus* 47.7 and 7.0 in the smaller set of LIS journals (Table 10).

In this much smaller set, the diversity in the citing dimension is significantly correlated to BC. In other words, citing behavior is more specific at the specialty level. The socio-cognitive structure of the field guides the variation. Table 10 shows the values of diversity in the case of *Scientometrics* at the three levels, respectively. Diversity is larger in the citing than cited dimension at the level of the full set. Limitation to the social sciences leads to losing citation in the citing dimension more than in the cited. As a consequence, diversity is larger in the cited than citing dimension at this level. Being at the edge of the LIS set, the journal cites more than it is cited by other journals in this set.

|  | N | Rao-Stirling | | $^2D^3$ | | Quantile values | | Σ cited | Σ citing | Σ Self-citation |
| --- | --- | --- | --- | --- | --- | --- | --- | --- | --- | --- |
|  |  | Cited | Citing | Cited | Citing | Cited | Citing |  |  |  |
| Complete Set | 11,359 | 0.35 | 0.43 | 3.39 | 6.84 | 45.0 | 97.0 | 5766 | 9158 | 1963 |
| Social science | 3,274 | 0.28 | 0.19 | 2.27 | 1.63 | 17.3 | 60.2 | 4570 | 6840 | 1963 |
| LIS | 86 | 0.16 | 0.25 | 1.45 | 1.98 | 7.0 | 47.7 | 3494 | 3676 | 1963 |

**Table 10**: *Scientometrics* at three levels of aggregation.

In sum, diversity is dependent on the delineations of the sample in which it is measured.

## 4. Decomposition of the diversity

In a next step we decompose the LIS set of 86 journals. Three journals (*Econtent*, *Restauror*, and *Z Bibl Bibl*) are not part of the large component, and therefore not included in this decomposition. Using VOSviewer, six groups are distinguished, of which one contains only a single journal (*Soc Sci Inform*). Figure 4 shows this map. Mean diversity values with standard errors for the five groups decomposed as sub-matrices are provided in Figure 5.

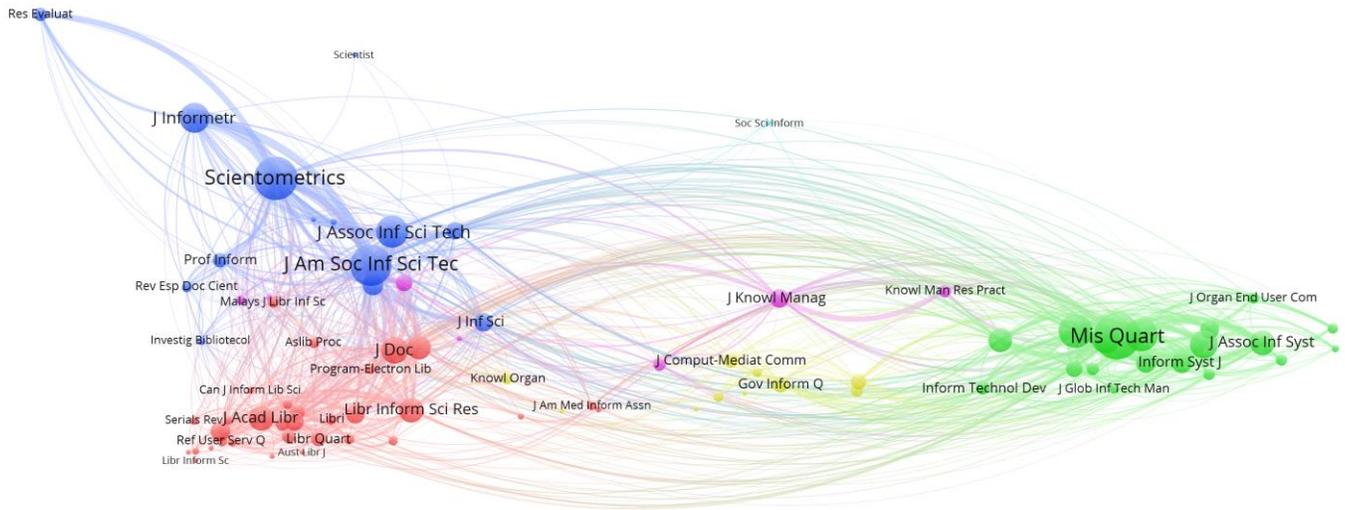
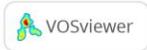

**Figure 4**: Clustering of the LIS set (n = 86) into five clusters using VOSviewer.

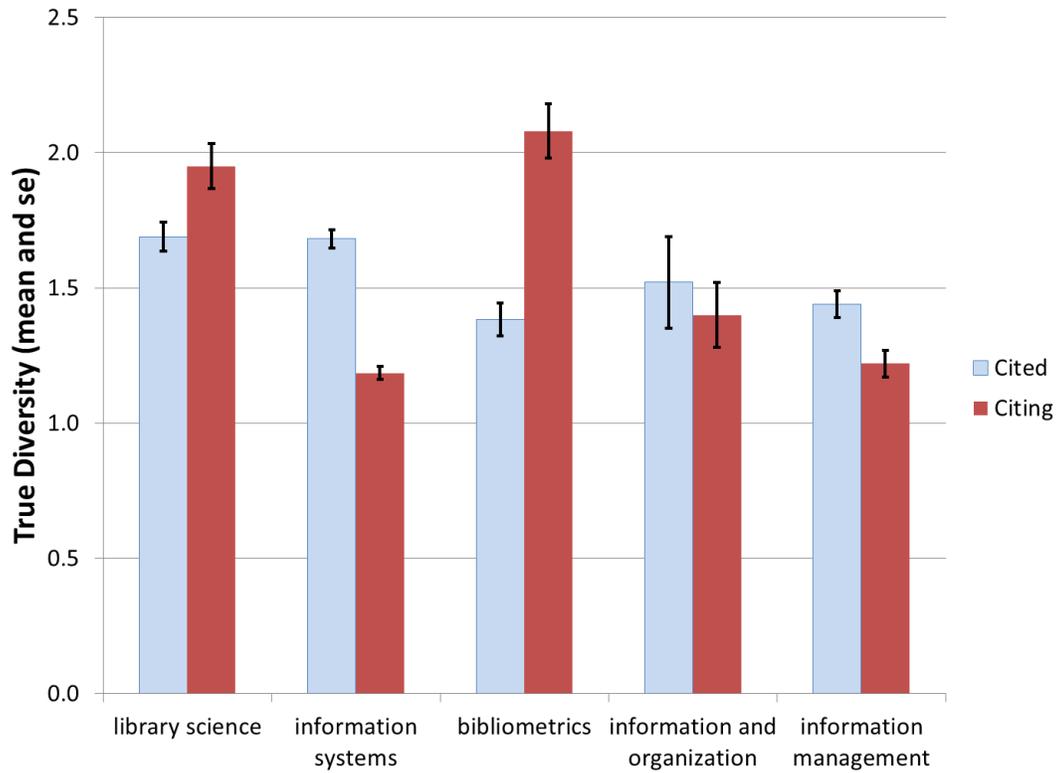

**Figure 5**: Average $^2D^3$ cited and citing for five subgroups of LIS journals (error bars with standard errors).

Based on the *post-hoc* Tukey test, two homogenous groups are distinguished in the citing dimension: library science and bibliometrics with relatively high citing scores, on the one side, and the other three with significantly lower scores, on the other. However, the distinction is not significant. In the cited direction, the set is statistically homogenous.

**Table 10**: Decomposition of the diversity in the LIS set.

| Specialism | $^2D^3$ cited | $\Delta$ cited | $^2D^3$ citing | $\Delta$ citing | N of journals |
|---|---|---|---|---|---|
| 1. library science | 54.02 | 12.36 | 59.11 | 13.19 | 32 |
| 2. information systems | 31.95 | 7.14 | 22.51 | 2.86 | 19 |
| 3. bibliometrics | 19.37 | 3.69 | 26.95 | 6.08 | 14 |
| 4. information and organization | 14.15 | 2.3 | 13.75 | 2.41 | 10 |
| 5. information management | 9.66 | 1.19 | 8.84 | 1.39 | 7 |
| Within-group | 129.15 | 26.68 | 131.16 | 25.93 | 82 |
| Total | 175.35 | 39.06 | 179.36 | 37.21 | 86 |

Within the subsets, however, the diversity scores are based on sub-matrices with corresponding cosine values. Table 11 provides the diversity values when all cell values are normalized in terms of the grand matrix. The difference between the total diversity and the sum of the within-group diversities is then by definition equal to the between-group diversity. Using the Tukey test with this design between-group diversity is significantly different from diversity in all the subsets with the exception of citing diversity in the subset of 19 journals labeled as "information science." Both cited and citing, "information science" and "library science" are considered as homogenous with between-group diversity. The other three specialism are considered as significantly different.

**Table 11**: Between-group diversity in the LIS set.

|  | $\Delta$ cited | $\Delta$ citing |
|---|---|---|
| 0. between-group | 20.82 | 21.61 |
|  | *53.3%* | *58.1%* |
| 1. library science | 8.33 | 8.21 |
| 2. information systems | 5.58 | 2.56 |
| 3. bibliometrics | 2.51 | 3.49 |
| 4. information and organization | 1.06 | 1.08 |
| 5. information management | 0.77 | 0.25 |
| **Total** | **39.06** | **37.21** |

Note that the total diversity is generated in a matrix of 82 times 81 or 6642 cells, whereas the within-group diversity is generated in subsets which add up to 1648 cells (24.8%).[10] In other words, diversity is concentrated in these groupings since they generate in 24.8% of the cells (100–53.3=) 46.7% and (100–58.1=) 41.9% of the total diversity in the cited and citing directions, respectively.

## 5. Diffusion and Integration

In the formula $\Delta = \sum_{\substack{i,j \\ (i \neq j)}} p_i p_j d_{ij}$ each combination of $i$ and $j$ provides a reference to two rows or columns in the other dimension. If along a row vector of being cited, for example as in Figure 1, $p_5 > 0$ and $p_6 > 0$, this means that the third and fourth citing units are bibliographically coupled by citing this document or set. In the other direction, the two units would be co-cited by a citing unit. Co-citation among diverse sources by a citing unit has been considered as integration (Wagner *et al.*, 2011), whereas bibliographic coupling can be considered conversely as diffusion (Rousseau *et al.*, 2017). Using the concepts of integration and diffusion of knowledge for citing and cited diversity, respectively, one can draw diffusion and integration networks by extracting the $k=1$ neighbourhoods; for example in Pajek. Figures 6A and 6B provide these networks for the journal *Scientometrics* in the LIS set (JCR 2015) as an example: 38 journals constitute the diffusion network (Fig. 6A) and 51 the knowledge integration network (Fig. 6B).

---

[10] (32 * 31) + (19 * 18) + (14 * 13) + (10 * 9) + (7 * 6) = 1648 cells.

**Figure 6A**: 38 journals in the knowledge diffusion network of *Scientometrics* in the LIS set 2015.

**Figure 6B**: 51 journals in the knowledge integration network of *Scientometrics* in the LIS set 2015.

For example, Figure 6B shows that the *Journal of the American Society for Information Science and Technology* plays a central role in the knowledge integration network. Articles in this journal are cited in *Scientometrics;* but only the *Journal of the Assocation for Information Science and Technology*—the current name of the same journal—is visible in the diffusion network (Figure 6A). Note that this analysis was pursued within the LIS set of 86 journals. Other journals outside the LIS field (e.g., *Research Policy*) are also important in the citation environment of *Scientometrics.*

**Conclusions and discussion**

Using journals as units of analysis, we addressed the question of whether interdisciplinarity can be measured in terms of betweenness centrality or diversity as indicators. We tried to pursue these ideas in considerable detail. It seems to us that the problem of measuring interdisciplinarity, however, remains unsolved because of the fluidity of the term "interdisciplinarity." The very concept means different things in policy discourse and in science studies. From a scientometric perspective, interdisciplinarity is difficult to define if there is no operational definition of the disciplines. The latter problem, however, has remained an unsolved problem in bibliometrics.

Bibliometricians often use the WoS Subject Categories as a proxy for disciplines, but these categories are pragmatic (e.g., Pudovkin & Garfield, 2002; Leydesdorff & Bornmann, 2016; Rafols & Leydesdorff, 2009). In this study, we build on the statistical decomposition of the JCR data using VOSviewer (Leydesdorff *et al*., 2017). The advantage of this approach is that the two problems—of decomposition and interdisciplinarity—are separated.

Our main conclusions are:

- The analysis at different levels of aggregation taught us that BC can be considered as a measure of multi-disciplinarity more than interdisciplinarity. Valued BC improves on binary BC because citation networks are valued. Marginal links should not be considered equal to central ones.
- Diversity in the *citing* dimension is very different (and statistically independent) from BC: it can also indicate non- or trans-disciplinarity. In local and applicational contexts, for example, the disciplinary origin of knowledge contributions may be irrelevant. In specialist contexts, however, citing diversity is coupled to the intellectual structures in the set(s) under study.
- Diversity in the *cited* dimension may come closest to an understanding of interdisciplinarity as a trade-off between structural selection and stochastic variation.
- Despite the absence of "balance"— the third element in Rafols & Meyer's (2010) definition of interdisciplinarity— Rao-Stirling "diversity" is often used as an indicator of interdisciplinarity; but it remains only an indicator of diversity.
- The co-citation of diverse contributions in a citing article has been considered as knowledge integration (Wagner *et al*., 2011; Rousseau *et al*., 2017). Analogously, but with the opposite direction in the arrows, diversity in bibliographic coupling can be considered as diffusion across domains. Using an example, we have demonstrated how these concepts can be elaborated into integration and diffusion networks.

- The sigma (Σ) in the formula (Eq. 2) makes it possible to distinguish between within-group and between-group diversity. In this respect, the diversity measure is as flexible as Shannon entropy measures (Theil, 1972). Differences in diversity can be tested for statistical significance using Bonferroni correction *ex post*. Homogenous and non-homogenous (sub)sets can thus be distinguished.

In other words, the problems of measurement could be solved to the extent that a general routine for generating diversity scores from networks is provided (see the Appendix). However, the interpretation of diversity as interdisciplinarity remains the problem. Diversity is very sensititive to the delineation of the sample; but is this also the case for interdisciplinarity? Is interdisciplinarity an intrinsic characteristic or can it only be defined (as more or less interdisciplinarity) in relation to a distribution?

We focused on journals in this study, but our arguments are not journal-specific. Some units of analysis, such as universities, are almost by definition multi-disciplinary or non-disciplinary. Non-disciplinarity can also be called "trans-disciplinary" (Gibbons *et al*., 1994). However, the semantic proliferation of Greek and Latin propositions—meta-disciplinary, epi-disciplinary, etc.—does not solve the problem of the operationalization of disciplinarity and then also interdisciplinarity.

In summary, we conclude that multi-disciplinarity is a clear concept that can be operationalized. Knowledge integration and diffusion refer to diversity, but not necessarily to interdisciplinarity. Diversity can flexibly be measured, but the score is dependent on the system of reference. We submit that a conceptualization in terms of variation and selection may prove more fruitful. For example, one can easily understand that variation is generated when different sources are cited, but to consider this variation as interdisciplinary knowledge integration is at best metaphorical.

Given this state of the art, policy analysts seeking measures to assess interdisciplinarity can be advised to specify first the relevant contexts, such as journal sets, comparable departments, etc. Networks in these environments can be evaluated in terms of BC and diversity. The routine provided in the appendix may serve for the latter purpose and any network analysis program can be used for measuring BC. (When the network can be measured at the interval scale, one is advised to use valued BC.) The arguments provided in this study may be helpful for the interpretation of the results; for example, by specifying methodological limitations.

**Acknowledgement**

We thank Wouter de Nooy for advice and are grateful to Thomson Reuters for JCR data.

**Appendix: A Routine for the Measurement of Diversity in Networks**

The routine net2rao.exe—available at http://www.leydesdorff.net/software/diversity/net2rao.exe —reads a network in the Pajek format (.net) and generates the files rao1.dbf and rao2.dbf. Rao1.dbf contains diversity values for each of the rows (named here "cited") and each of the columns (named "citing"). Rao2.dbf is needed for the computation of cell values (see here below).

The input file is preferentially saved by Pajek so that the format is consistent. Use the standard edge-format. The user is first prompted for the name of this .net-file. The output contains the values of both Rao-Stirling diversity and so-called "true" diversity (labels: "Zhang_ting" and "Zhang_ted"; see Zhang *et al*., 2016; cf. Jost, 2006).

By changing the default "No" into "Yes," one can make the program write two files, labeled res_ting and res_ted, containing detailed information for each pass. These files can be used for detailed decompositions. However, the files may grow rapidly in size (> 1 GB). All files are overwritten in later runs; one is advised to save them under other names or in other folders.